\documentclass[sigconf]{acmart}

\AtBeginDocument{%
  \providecommand\BibTeX{{%
    \normalfont B\kern-0.5em{\scshape i\kern-0.25em b}\kern-0.8em\TeX}}}

\copyrightyear{2023}
\acmYear{2023}
\setcopyright{licensedothergov}\acmConference[KDD '23]{Proceedings of the 29th ACM SIGKDD Conference on Knowledge Discovery and Data Mining}{August 6--10, 2023}{Long Beach, CA, USA}
\acmBooktitle{Proceedings of the 29th ACM SIGKDD Conference on Knowledge Discovery and Data Mining (KDD '23), August 6--10, 2023, Long Beach, CA, USA}
\acmPrice{15.00}
\acmDOI{10.1145/3580305.3599507}
\acmISBN{979-8-4007-0103-0/23/08}

\usepackage{xcolor}
\usepackage{amsmath}
\usepackage{diagbox}
\usepackage{tikz}
\usetikzlibrary{arrows}
\usepackage{graphicx}
\usetikzlibrary{shapes}
\usetikzlibrary{datavisualization,arrows.meta}
\usepackage{pgfplots}
\usepackage{multirow}
\usepackage{algorithm}
\usepackage{algpseudocode}\usepackage{caption}
\usepackage{amsfonts}
\usepackage{nicematrix}
\usepackage{pifont}
\newcommand{\cmark}{\ding{51}}%
\newcommand{\xmark}{\ding{55}}%

\usepackage{amsmath,amsfonts,bm}

\def\eqref#1{equation~\ref{#1}}

\def\1{\bm{1}}

\DeclareMathAlphabet{\mathsfit}{\encodingdefault}{\sfdefault}{m}{sl}
\SetMathAlphabet{\mathsfit}{bold}{\encodingdefault}{\sfdefault}{bx}{n}

\usepackage{hyperref}
\usepackage{url}
\usepackage{booktabs}
\usepackage{graphicx}
\usepackage{bbm}
\usepackage{caption}
\usepackage{subcaption}
\usepackage{balance} 

\begin{document}
\title{Source-Free Domain Adaptation with Temporal Imputation for Time Series Data}


\author{Mohamed Ragab}
\orcid{0000-0002-2138-4395}

\affiliation{%
\institution{Institute for Infocomm Research, Agency for Science Technology and Research (A*STAR)}
\streetaddress{1 Fusionopolis Way}
\country{Singapore}
\postcode{138632}}
\affiliation{
  \institution{Centre for Frontier AI Research, Agency for Science Technology and Research (A*STAR)}
  \streetaddress{1 Fusionopolis Way}
  \country{Singapore}
  \postcode{138632}
}
\email{mohamedr002@e.ntu.edu.sg}

\author{Emadeldeen Eldele}
\orcid{0000-0002-9282-0991}
\affiliation{
  \institution{Nanyang Technological University}
  \streetaddress{50 Nanyang Ave}
  \country{Singapore}
  \postcode{639798}
}
\affiliation{%
\institution{Centre for Frontier AI Research,  Agency for Science Technology and Research (A*STAR)}
\streetaddress{1 Fusionopolis Way}
\country{Singapore}
\postcode{138632}}
\email{emad0002@ntu.edu.sg}

\author{Min Wu}
\affiliation{%
  \institution{Institute for Infocomm Research, Agency for Science Technology and Research (A*STAR)}
 \streetaddress{1 Fusionopolis Way}
  \country{Singapore}
  \postcode{138632}}
\email{wumin@i2r.a-star.edu.sg}

\author{Chuan-Sheng Foo}

\affiliation{%
\institution{Institute for Infocomm Research, Agency for Science Technology and Research (A*STAR)}
\streetaddress{1 Fusionopolis Way}
\country{Singapore}
\postcode{138632}}
\affiliation{
  \institution{Centre for Frontier AI Research, Agency for Science Technology and Research (A*STAR)}
  \streetaddress{1 Fusionopolis Way}
  \country{Singapore}
  \postcode{138632}
}
\email{foo_chuan_sheng@i2r.a-star.edu.sg}

\author{Xiaoli Li}

\affiliation{%
\institution{Institute for Infocomm Research, Agency for Science Technology and Research (A*STAR)}
\streetaddress{1 Fusionopolis Way}
\country{Singapore}
\postcode{138632}}
\affiliation{
  \institution{Centre for Frontier AI Research, Agency for Science Technology and Research (A*STAR)}
  \streetaddress{1 Fusionopolis Way}
  \country{Singapore}
  \postcode{138632}
}
\email{xlli@i2r.a-star.edu.sg}

\author{Zhenghua Chen}
\authornote{Corresponding Author}
\affiliation{%
\institution{Institute for Infocomm Research, Agency for Science Technology and Research (A*STAR)}
\streetaddress{1 Fusionopolis Way}
\country{Singapore}
\postcode{138632}}
\affiliation{
  \institution{Centre for Frontier AI Research, Agency for Science Technology and Research (A*STAR)}
  \streetaddress{1 Fusionopolis Way}
  \country{Singapore}
  \postcode{138632}
}
\email{chen0832@e.ntu.edu.sg}
\renewcommand{\shortauthors}{Mohamed Ragab et al.}

\begin{abstract}
Source-free domain adaptation (SFDA) aims to adapt a pretrained model from a labeled source domain to an unlabeled target domain without access to the source domain data, preserving source domain privacy. Despite its prevalence in visual applications, SFDA is largely unexplored in time series applications. The existing SFDA methods that are mainly designed for visual applications may fail to handle the temporal dynamics in time series, leading to impaired adaptation performance. To address this challenge, this paper presents a simple yet effective approach  for source-free domain adaptation on time series data, namely MAsk and imPUte (MAPU). First, to capture temporal information of the source domain, our method performs random masking on the time series signals while leveraging a novel temporal imputer to recover the original signal from a masked version in the embedding space. Second, in the adaptation step, the imputer network is leveraged to guide the target model to produce target features that are temporally consistent with the source features. To this end, our MAPU can explicitly account for temporal dependency during the adaptation while avoiding the imputation in the noisy input space. Our method is the first to handle temporal consistency in SFDA for time series data and can be seamlessly equipped with other existing SFDA methods. Extensive experiments conducted on three real-world time series datasets demonstrate that our MAPU achieves significant performance gain over existing methods. Our code is available at \url{https://github.com/mohamedr002/MAPU_SFDA_TS}.
\end{abstract}


\maketitle

\begin{figure}
    \centering
    \includegraphics[width = 0.45\textwidth]{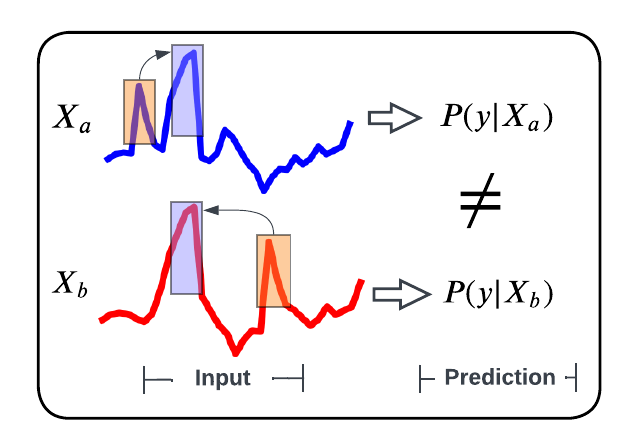}
    \caption{How do the temporal relations matter in time series? Despite similarities in the values of the two signals, variations in the temporal position of their observations can result in different predictions.}
    \label{fig:t_shift}
\end{figure}

\begin{figure*}
    \centering
    \includegraphics[width = 0.8\textwidth]{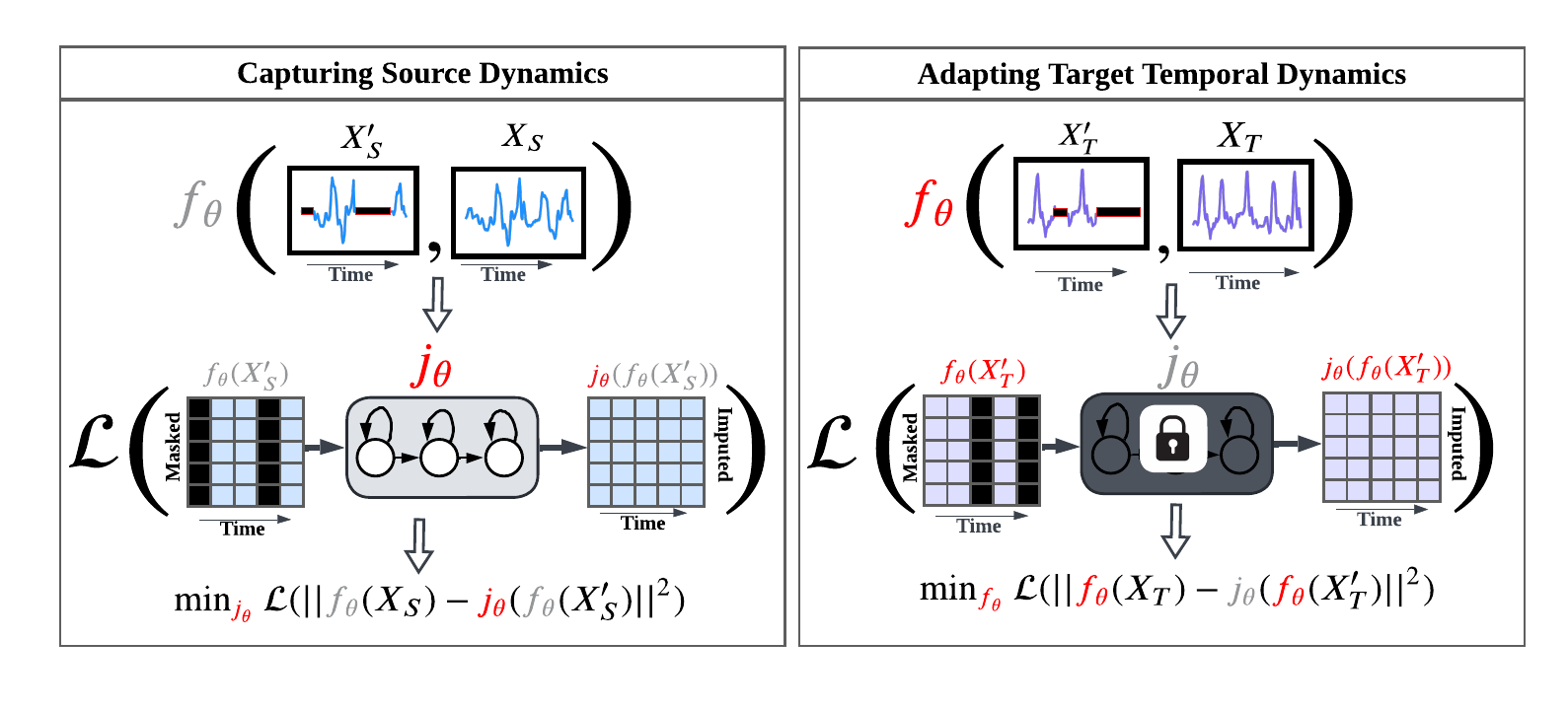}
    \caption{Adaptation with Temporal Imputation. \underline{Left}: A temporal imputer network is trained to predict the full sequence from its masked version to capture the temporal information of the source domain. \underline{Right}: Once trained, the temporal imputer network guides the target model to produce features that are temporally consistent with the source domain. (Best in viewed in colors. Components in red color are trainable, while those in gray color are non-trainable).}
    \label{fig:key_idea}
\end{figure*}

\section{Introduction}
Deep learning has achieved impressive performance in numerous time series applications, such as machine health monitoring, human activity recognition, and healthcare. However, this success heavily relies on the laborious annotation of large amounts of data. To address this issue, unsupervised domain adaptation (UDA) has gained traction as a way to leverage pre-labeled source data for training on unlabeled target data, while also addressing the distribution shift between the two domains~\cite{da_survey}. There is a growing interest in applying UDA to time series data~\cite{adatime}, with existing methods seeking to minimize statistical distance across the source and target features \cite{dsan,aaai_ts_uda} or using adversarial training to find domain invariant features \cite{codats,calda,ts_da_attn,slarda}. However, these approaches require access to the source data during the adaptation process, which may not be always possible, due to data privacy regulations. 

To address this limitation, a more practical setting, i.e., source-free domain adaptation (SFDA), has been proposed, where only a source-pretrained model is available during the adaptation process~\cite{sfda_kim}. In recent years, several SFDA methods have been developed for visual applications \cite{shot, sfda_kim,ss_sfda,sahoo_sfda,cpga}. One prevalent paradigm has incorporated some auxiliary tasks to exploit the characteristics of visual data to improve the source-free adaptation \cite{SHOT++,sl-sfda,ss_sfda}. 
However, all these methods are primarily designed for visual applications and may fail to handle the temporal dynamics of time series data.

In time series data, temporal dependency refers to the interdependence between values at different time points, which has a significant impact on predictions \cite{temp_shift_1}. As demonstrated in Figure \ref{fig:t_shift}, even two signals with similar observations can lead to differing predictions if the temporal order is different. Such temporal dynamics make adapting the temporal information between two shifted domains a key challenge in unsupervised domain adaptation. The problem becomes even more challenging under source-free adaptation settings, where no access to the source data is provided during the target adaptation. Therefore, our key question is how to effectively adapt temporal information in time series data in the absence of the source data.

In this work, we address the above challenges and propose a novel SFDA approach, i.e., MAsk and imPUte (MAPU), for time series data. Our method trains an autoregressive model to capture the temporal information on the source domain, which is then transferred to the target domain for adaptation. 
The key steps of our approach are illustrated in Figure \ref{fig:key_idea}. First, the input signal undergoes temporal masking. Both the masked signal and the original signal are then fed into an encoder network, which generates the corresponding feature representation. Subsequently, the temporal imputation network is trained to impute the original signal from the masked signal in the feature space, enabling smoother optimization for the temporal imputation task. During adaptation, the imputation network is used to guide the target model to generate target features that can be imputed by the source imputation network. Our method is versatile and can be integrated with the existing SFDA methods to provide them with temporal adaptation capability.
The main contributions of this work can be summarized as follows: 
\begin{itemize}
    \item To the best of our knowledge, we are the first to achieve the source-free domain adaptation for time series applications.

    \item We propose a novel temporal imputation task to ensure sequence consistency between the source and target domains. 

    \item We propose a versatile methodology for integrating temporal adaptation capability into existing SFDA methods.

    \item We conduct extensive experiments and demonstrate that our approach results in a significant improvement in adaptation performance on real-world data, and is particularly effective for time series adaptation tasks.
\end{itemize}

\section{Related work}

\subsection{Time series Domain Adaptation}
Several methods have been proposed to address the challenge of distribution shift in time series data. These methods can be broadly categorized into two groups: discrepancy-based methods and adversarial-based methods. Discrepancy-based methods use statistical distances to align the feature representations of the source and target domains. For instance, AdvSKM leverages the maximum mean discrepancy (MMD) distance in combination with a hybrid spectral kernel to consider temporal dependencies during domain adaptation \cite{dskn}. Another example is SASA, which learns the association structure of time series data to align the source and target domains \cite{aaai_ts_uda}. On the contrary, adversarial-based methods use adversarial training to mitigate the distribution shift between the source and target domains. For instance, CoDATS utilizes a gradient reversal layer (GRL) for adversarial training with weak supervision on multi-source human activity recognition data \cite{codats}. Furthermore, DA\_ATTN couples adversarial training with an un-shared attention mechanism to preserve the domain-specific information \cite{ts_da_attn}. Recently, SLARDA presents an autoregressive adversarial training approach for aligning temporal dynamics across domains \cite{slarda}. 

Albeit promising, the design of these methods is based on the assumption that source data is available during the adaptation step. However, accessing source data may not be possible in practical situations due to privacy concerns or storage limitations. Differently, our MAPU adapts a model pretrained on source data to new domains without access to source data during adaptation, which can be a more practical solution for high-stake applications. 

\begin{figure*}
    \centering
    \includegraphics[width = 0.75\textwidth]{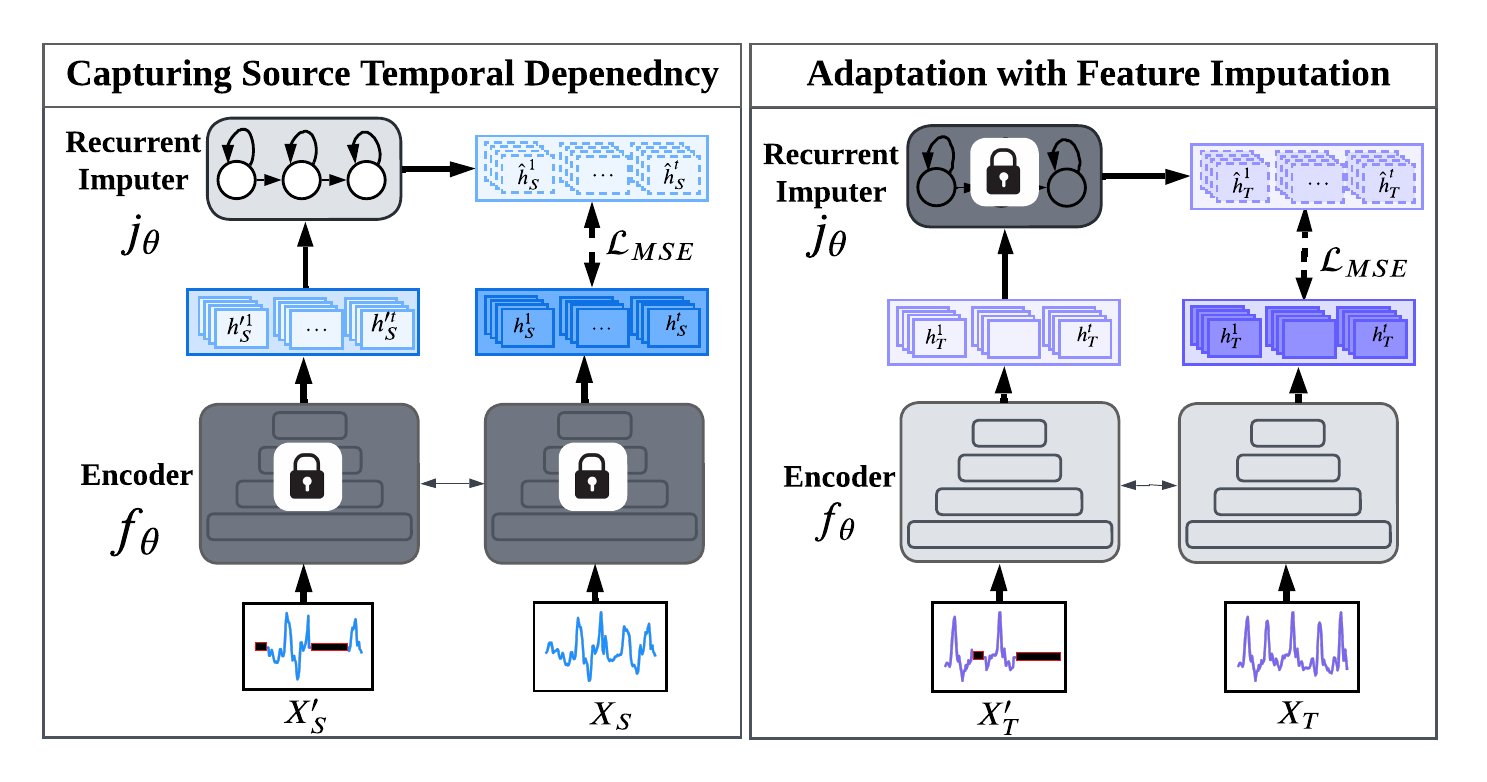}
    \caption{Adaptation with Temporal Imputation for time series data. \underline{Left}: The pretraining stage of the temporal imputer network $j_\theta$ to capture the temporal dynamics of the source domain. First, we perform random masking across the time dimension of the source signal. Given the original source signal $X_S$ and its temporally masked signal $X_S^\prime$, the encoder network $f_\theta$ is used to generate the corresponding latent features $H_S$ and $H_S^\prime$ respectively. Subsequently, $j_\theta$ is updated to produce imputed features $\hat{H}_S$ from masked features $H_S^\prime$ using the mean square error loss. \underline{Right}: The adaptation stage of the encoder network on the target domain data. The encoder $f_\theta$ is updated to produce source-like features that are imputable by the pretrained $j_\theta$.}
    \label{fig:temp_adapt}
\end{figure*}

\subsection{Source Free Domain Adaptation}  
Source-Free Domain Adaptation (SFDA) is a new problem-setting, where we do not have access to the source domain data during adaptation. This objective can be achieved in several ways. One approach is to leverage a model pretrained on the source domain to generate synthetic source-like data during the adaptation step \cite{cpga,3c_gan,sdda,sahoo_sfda}. Another approach is to use adversarial training between multiple classifiers to generalize well to the target classes \cite{d_mcd,a2_net}. Another prevalent approach uses softmax scores or their corresponding entropy to prioritize confident samples for pseudo-labeling, assuming that the model should be more confident on source samples and less confident on target samples \cite{shot,sfda_kim,ss_sfda}.

Despite the strong potential demonstrated by these methods, they are primarily designed for visual applications and may fail to effectively align temporal dynamics in time series data. In contrast, our method addresses this challenge through a novel temporal imputation task, ensuring temporal consistency between domains during the adaptation.

\section{Methodology}
\subsection{Problem definition} 
Given a labeled source domain $\mathcal{D}_S = \{{X}_S^i, {y}_S^i\}_{i=1}^{n_S} $, where ${X}_S \in \mathcal{X}_S$ can be a uni-variate or multi-variate time series data with a sequence length $L$, while $y_S \in \mathcal{Y}_S$ represents the corresponding labels. In addition, we have an unlabeled target domain $\mathcal{D}_T= \{{X}_T^i\}_{i=1}^{n_T} $, where ${X}_T \in \mathcal{X}_T$, and it also shares the same label space with $\mathcal{D}_S$. Following the existing UDA settings, we assume a difference across the marginal distributions, i.e., $P(X_S) \neq P(X_T)$, while the conditional distributions are stable, i.e., $P(y_S|X_S) \approx P(y_T|X_T)$. 

This work aims to address the source-free domain adaptation problem, where access to source data is strictly prohibited during the adaptation phase to ensure data privacy.  Furthermore, we adopt the vendor-client source-free paradigm \cite{vc1,vc2,vc3,vc4}, which allows the influence of the source pretraining stage. This assumption is realistic in use cases where there is a collaboration between various entities, but it is not possible to share source data due to data privacy, security, or regulatory issues.

\subsection{Overview}
\begin{figure*}[h]
    \centering
    \includegraphics[width=\textwidth]{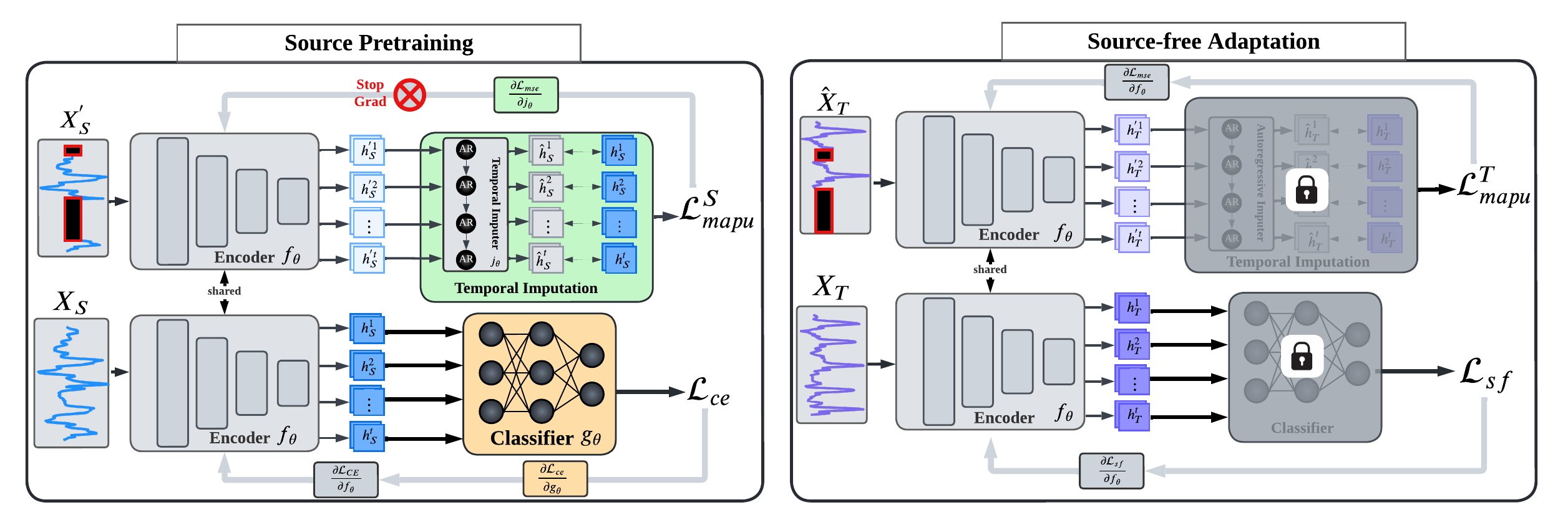}
    \caption{Integrating our temporal imputation with other source-free methods (Best viewed in colors). \underline{Left}: In the pretraining stage, the source model is trained using conventional cross-entropy loss $\mathcal{L}_{ce}$, and the temporal imputer network is trained using $\mathcal{L}_{mapu}^S$ to impute the features of the masked signals and capture the  source temporal information on the feature space. \underline{Right}: In the adaptation stage, the target model is jointly trained with both a generic source-free adaptation loss $\mathcal{L}_{sf}$ and our temporal imputation loss $\mathcal{L}_{mapu}^T$ to perform the adaptation while ensuring temporal consistency with the source features.}
    \label{fig:overall}
\end{figure*}

We present our MAPU to achieve source-free adaptation on time series temporal data while considering the temporal dependencies across domains. The pipeline of the proposed method is illustrated in Figure~\ref{fig:overall}. Given the input signal and a temporally masking signal, our method comprises two stages: (1) training an autoregressive network, referred to as the \textit{imputer network}, which captures the temporal information of the source domain through a novel temporal imputation task, and (2) leveraging the source-pretrained imputer network to guide the target encoder towards producing temporally consistent target features in the adaptation stage. Next, we will first elaborate on the temporal masking procedure before delving into the details of each stage.

\subsection{Temporal Masking}
In this section, we explain our process of temporal masking. We start by dividing the input signal, $X$, into several blocks along the time dimension. Then, we randomly choose some of these blocks and set their values to zero, creating a masked version of the signal called $X^\prime$. This process is applied to both the source and target domains. Our aim is to challenge the model to use the information from surrounding blocks to fill in the missing parts and capture the temporal dependencies in the input signal. Further discussion on the impact of the masking ratio on the adaptation performance can be found in the experiment section.
\subsection{Capturing Source Temporal Dynamics}
In the pretraining stage, current methods typically map the source data from the input space to the feature space using an encoder network, represented as $f_\theta: \mathcal{X}_S \rightarrow \mathcal{H}_S$. The extracted features are then passed through a classifier network, $g_\theta:\mathcal{H}_S \rightarrow \mathcal{Y}_S$, to make class predictions for the source data. 
However, to effectively adapt to other time series domains, it is important to consider the temporal relations in the source domain. Using only cross-entropy for training the source network may neglect this aspect. To address this, we propose a temporal imputation task that aims to recover the input signal from a temporally masked signal in the feature space.

The imputation task is performed by an imputer network $j_\theta$ that takes the masked signal and maps it to the original signal. The input signal $X_S$ and masked signal $X_S^\prime$ are first transformed into their corresponding feature representations $H_S$ and $H_S^\prime$ by the encoder $f_\theta$. The task of the imputer network is represented as $\hat{H}_S^\prime = j_\theta(f_\theta(X_S^{\prime})) \rightarrow H_S = f_\theta(X_S)$, where $\hat{H}_S^\prime$ is the imputed signal. The imputer network is trained to minimize the mean square error between the features of the original signal and the imputed signal, which can be formulated as:
\begin{align}
    \min_{j_\theta}\mathcal{L}_{mapu}^S = \frac{1}{n}\sum_{i=1}^{n_S}\left\|f_\theta(X_S^i) - j_\theta(f_\theta(X_S^{\prime i}))\right\|_2^2,
\end{align}
where $H_S = f_\theta(X_S^i)$ are the latent features of the original signal, $\hat{H}_S = j_\theta(f_\theta(X_S^{\prime i}))$ is the output of the imputer network, and $n_S$ is the total number of source samples.

\subsection{Temporal Adaptation with Feature Imputation}
In the adaptation stage, the goal is to train the target encoder network to produce target features temporally consistent with the source features. The target encoder network $f_\theta$ is used to extract latent feature representations from a target sample $X_T$ and its masked version $X_T^\prime$. The fixed source-pretrained imputer network $j_\theta$ is then used to reconstruct the features of the original signal from the masked features. However, due to domain differences, the source imputer may not be able to accurately reconstruct the target features. Thus, the encoder network $f_\theta$ is updated to produce target features that can be accurately reconstructed by the imputer network. This can be expressed as the following optimization problem:

\begin{align}
\min_{f_\theta}\mathcal{L}_{mapu}^T = \frac{1}{n}\sum_{i=1}^{n_T}\left\|f_\theta(X_T)^i - j_\theta(f_\theta(X_T^\prime)^i))\right\|_2^2,
\end{align}
where $H_T= f_\theta(X_T)$ are the original target features, $\hat{H}_T = j_\theta(f_\theta(X_T^\prime))$ are the adapted target features produced by the imputer network to minimize the mean square error loss, and $n_T$ is the total number of  target samples. Notably, only the encoder network is optimized, producing features that can be accurately imputed by the fixed source-pretrained imputer network. To reduce the imputation loss, the adapted target features should be temporally consistent with the source features. 

Algorithm~1 illustrates the adaptation procedure via temporal imputation. The process starts by first constructing a temporally masked version of the input target sample represented as $X_T^{\prime}$. Next, the source-pretrained encoder is used to extract the latent features of both the original signal and the temporally masked signal, represented as $H_T$ and $H_T^{\prime}$, respectively. Finally, the encoder network is updated to make features of the masked signal recoverable by the source-pretrained imputer network, using the mean square error loss in Equation 2.

\begin{algorithm}[tb]
\caption{Adaptation with Temporal Imputation}
   \label{alg:example}
\begin{algorithmic}[1]
\State \textbf{Input:} Target sample $X_T$, source pretrained encoder $f_\theta$, source classifier $g_\theta$, imputer network $j_\theta$ 
\State \textbf{Initialize:} Construct temporally masked version of $X_T$: $X_T^{\prime}$
\State \textbf{Extract:} Latent feature representations: $H_T = f_\theta(X_T)$, $H_T^{\prime} = f_\theta(X_T^{\prime})$
\State \textbf{Impute:} Masked features using imputer network: $\hat{H}_T = j_\theta(H_T^{\prime})$
\State \textbf{Update:} Encoder $f_\theta$ to produce target features that can be accurately reconstructed by $j_\theta$ using Equation~2
\State \textbf{Output:} Updated encoder $f_\theta$ 
\end{algorithmic}
\end{algorithm}

\subsection{Integration with Other Source-free Methods}
Our proposed MAPU is generic and can be integrated with other source-free adaptation methods. Typically, source-free adaptation involves a two-stage training procedure: (1) pretraining the source model with source domain data, and (2) adapting the pretrained model to the target domain. As shown in Figure~\ref{fig:overall}, our MAPU can be seamlessly integrated into existing SFDA methods in both stages.

In the pretraining stage, MAPU operates in the feature space by training the temporal imputation network, $j_\theta$, to capture the temporal information from the source domain. The loss associated with the temporal imputation task does not propagate to the encoder model, $f_\theta$. As a result, the encoder can be trained exclusively with the conventional cross-entropy loss, ensuring that the imputation task does not negatively impact the pretraining performance. The total pretraining loss is formalized as:

\begin{align}
&\min_{f_\theta, j_\theta}
\mathcal{L}_{\mathrm{S}}= -\mathbb{E}_{({X}_S,y_S) \sim \mathcal{D}_S} \mathcal{L}_{\mathrm{ce}} + \mathcal{L}_{\mathrm{mapu}}^S,
\end{align}
where $\mathcal{L}_{\mathrm{ce}} = \sum_{k=1}^K \mathbbm{1}_{[y_S = k]} \log(\hat{p}_k)$ represents the standard cross-entropy loss between the predicted label and the true label, $\hat{p}k^i$ represents the predicted probability for class $k$ and sample $i$, and $\mathcal{L}_{\mathrm{mapu}}^S$ represents the training loss for our temporal imputer network on the source data to capture the source temporal information.

In the target adaptation step, the objective is to optimize the target encoder, $f_\theta$, by balancing the temporal imputation loss and the generic source-free loss to achieve temporal consistency and perform adaptation on the target domain. This can be formalized as follows:

\begin{align}
\min_{f_\theta} L_{T} = \mathbb{E}_{{X}_T \sim \mathcal{D}_T} \mathcal{L}_{sf} + \alpha \mathcal{L}_{mapu}^T,
\end{align}
where $\alpha$ is a hyperparameter that regulates the relative importance of the temporal imputation task, and $\mathcal{L}_{sf}$ represents the generic loss used by the SFDA method to adapt the target domain to the source domain.

\section{Experimental Settings}
\subsection{Datasets}
We evaluate our proposed method on three real-world datasets spanning three time series applications, i.e., machine fault diagnosis, human activity recognition, and sleep stage classification. The selected datasets differ in many aspects, as illustrated in Table \ref{tbl:datasets},  which leads to a considerable domain shift across different domains.

\subsubsection{UCIHAR Dataset} 
This dataset focuses on human activity recognition tasks. Three types of sensors have been used to collect the data, i.e., accelerator sensor, gyroscope sensor, and body sensor, where each sensor provides three-dimensional readings, leading to a total of 9 channels per sample, with each sample containing 128 data points. The data is collected from 30 different users and each user is considered as one domain. In our experiments, five cross-user experiments are conducted, where the model is trained on one user and tested on different users to evaluate its cross-domain performance \cite{uciHAR_dataset}.

\subsubsection{Sleep Stage Classification (SSC) Dataset} 
The Sleep Stage Classification (SSC) task involves categorizing Electroencephalography (EEG) signals into five distinct stages, namely Wake (W), Non-Rapid Eye Movement stages (N1, N2, N3), and Rapid Eye Movement (REM). To accomplish this, we utilize the Sleep-EDF dataset \cite{sleepEDF_dataset}, which comprises EEG readings from 20 healthy subjects. In line with previous studies \cite{attnSleep_paper}, we select a single channel, specifically Fpz-Cz, and utilize 10 subjects to construct five cross-domain experiments.

\subsubsection{Machine Fault Diagnosis (MFD) Dataset}
This dataset has been collected by Paderborn university for the fault diagnosis application, where the vibration signals are leveraged to identify different types of incipient faults. The data has been collected under four different working conditions. Each data sample consists of a single univariate channel and 5120 data points following previous works \cite{fd_dataset,slarda}. In our experiments, each working condition is considered as one domain, where we utilize five different cross-condition scenarios to evaluate the domain adaptation performance. 

More details about the datasets are included in Table~\ref{tbl:datasets}.

\begin{table}[!tbh]
\centering
\caption{Details of the adopted datasets (C: \#channels, K: \#classes, L: sample length).}
\resizebox{\columnwidth}{!}{
\begin{NiceTabular}{l|ccc|cc}
\toprule
\textbf{Dataset} & \textbf{C}  & \textbf{K} & \textbf{L} & \# training samples & \# testing samples \\ \midrule

UCIHAR &  9 & 6 & 128 & 2300 & 990 \\ 
SSC & 1 & 5 & 3000 & 14280 & 6130 \\ 
MFD & 1 & 3 & 5120 & 7312 & 3604 \\ 
\bottomrule
\end{NiceTabular}
}
\label{tbl:datasets}
\end{table}
\begin{table*}[h]
\centering
\caption{Detailed results of the five UCIHAR cross-domain scenarios in terms of MF1 score.}
\begin{NiceTabular}{@{}l|c|ccccc|c@{}} 
\toprule 
Algorithm & SF & 2$\rightarrow$11 & 12$\rightarrow$16 & 9$\rightarrow$18 & 6$\rightarrow$23 & 7$\rightarrow$13 & AVG\\ \midrule
DDC & \xmark & 60.0$\pm$13.32 & 66.77$\pm$8.46 & 61.41$\pm$5.80 & 88.55$\pm$1.42 & 77.29$\pm$2.11 &   75.67 \\
DCoral &\xmark &  67.2$\pm$13.67 & 64.58$\pm$8.72 & 54.38$\pm$9.69 & 89.66$\pm$2.54 & 90.46$\pm$2.96 &   77.71 \\
HoMM &\xmark &  83.54$\pm$2.99 & 63.45$\pm$2.07 & 71.25$\pm$4.42 & 94.97$\pm$2.49 & 91.41$\pm$1.33 &   84.10 \\

MMDA &\xmark &  72.91$\pm$2.78 & \textbf{74.64$\pm$2.88}& 62.62$\pm$2.63 & 91.14$\pm$0.46 & 90.61$\pm$2.00 &  81.40  \\


DANN &\xmark &  98.09$\pm$1.68 & 62.08$\pm$1.69 & 70.7$\pm$11.36 & 85.6$\pm$15.71 & \underline{93.33$\pm$0.00} &   84.97 \\

CDAN &\xmark &  \underline{98.19$\pm$1.57} & 61.20$\pm$3.27 & 71.3$\pm$14.64 & {96.73$\pm$0.00} & \underline{93.33$\pm$0.00}  &   \underline{86.79} \\


CoDATS &\xmark &  86.65$\pm$4.28 & 61.03$\pm$2.33 & \underline{80.51$\pm$8.47} & 92.08$\pm$4.39 & 92.61$\pm$0.51  &   85.47 \\

AdvSKM  &\xmark &  65.74$\pm$2.69 & 60.52$\pm$1.99 & 53.25$\pm$5.19 & 79.63$\pm$8.52 & 88.89$\pm$3.12 &   74.67 \\ \midrule

SHOT & \cmark
&\textbf{100.0$\pm$0.00} 
& \underline{70.76$\pm$6.22} 
& 70.19$\pm$8.99	
& \textbf{98.91$\pm$1.89} 
& 93.01$\pm$0.57	
& 86.57 \\ 
NRC & \cmark 
& 97.02$\pm$2.82 
&72.18$\pm$0.59 
& 63.10$\pm$4.84 
& 96.41$\pm$1.33
& 89.13$\pm$0.54 
& 83.57   \\ 
AaD  & \cmark
& 98.51$\pm$2.58
& 66.15$\pm$6.15 
& 68.33$\pm$11.9
& \underline{98.07$\pm$1.71}
& 89.41$\pm$2.86 
& 84.09  \\ \midrule
\textbf{MAPU}   & \cmark 
& \textbf{100.0$\pm$0.00}  
&  67.96$\pm$4.62 
& \textbf{82.77$\pm$2.54} 
& 97.82$\pm$1.89 
& \textbf{99.29$\pm$1.22} 
& \textbf{89.57 }\\ 
 \bottomrule
\end{NiceTabular}
\label{table:har}
\end{table*}




\subsection{Implementation Details}
\paragraph{Encoder Design}
In our study, we adopt the encoder architecture presented in existing works \cite{adatime,ts_tcc}, which is a 1-dimensional convolutional neural network composed of three layers with filter sizes of 64, 128, and 128 respectively. Each conventional layer was followed by the application of a rectified linear unit activation function and batch normalization. 
\paragraph{MAPU Parameters}
For the purpose of temporal masking, a masking ratio of 1/8 is utilized across all datasets in our experiments. To perform the imputation task, a single-layer recurrent neural network with a hidden dimension of 128 is employed for all datasets. In addition, our method includes a primary hyperparameter, $\alpha$, which is set to 0.5 for all datasets in our evaluation. 
\paragraph{Unified Training Scheme}
To provide a fair and valid comparison with source-free baseline methods, we adhered to their established implementations \cite{shot,nrc,aad} while incorporating the same backbone network and training procedures utilized in our proposed method. 
In accordance with the AdaTime framework \cite{adatime}, all the models are trained for a total of 40 epochs, using a batch size of 32, with a learning rate of 1e-3 for UCIHAR and 1e-4 for SSC and MFD. 
Also, the macro F1-score (MF1) metric \cite{adatime} has been used to ensure a reliable evaluation under data imbalance situations, where we report the mean and the standard deviation of three consecutive runs for each cross-domain scenario.

\subsection{Baseline Methods}
To evaluate the performance of our model, we compare it against conventional UDA approaches that assume access to source data during adaptation. These baselines are adapted from the AdaTime benchmark \cite{adatime}. Additionally, we compare our model against recent source-free domain adaptation methods. To ensure fair evaluation, we re-implement all source-free baselines in our framework, while ensuring the same backbone network and training schemes. Overall the compared methods are as follows: 
\paragraph{Conventional UDA methods}
\begin{itemize}
    \item Deep Domain Confusion (DDC) \cite{ddc}: leverages the MMD distance to align the source and target features.
    \item Deep Correlation Alignment  (DCORAL) \cite{deep_coral}: aligns the second-order statistics of the source and target distributions in order to effectively minimize the shift between the two domains. 
    \item High-order Maximum Mean Discrepancy (HoMM) \cite{HoMM}: aligns the high-order moments to effectively tackle the discrepancy between the two domains. 
    \item Minimum Discrepancy Estimation for Deep Domain Adaptation (MMDA) \cite{MMDA}: combines the MMD and correlation alignment with entropy minimization to effectively address the domain shift issue.
    \item Domain-Adversarial Training of Neural Networks (DANN) \cite{DANN}: leverages gradient reversal layer to adversarially train a domain discriminator network against an encoder network.
    \item Conditional Domain Adversarial Network (CDAN) \cite{CDAN}: realizes a conditional adversarial alignment by integrating task-specific knowledge with the features during the alignment step for the different domains.
    \item Convolutional deep adaptation for time series (CoDATS) \cite{codats}: employs adversarial training with weak supervision to enhance the adaptation performance on time series data.
    \item Adversarial spectral kernel matching (AdvSKM)\cite{dskn}: introduces adversarial spectral kernel matching to tackle the challenges of non-stationarity and non-monotonicity present in time series data.
\end{itemize}

\paragraph{Source-free methods}

\begin{itemize}
    \item Source Hypothesis Transfer (SHOT) \cite{shot}: minimizes information maximization loss with self-supervised pseudo labels to identify target features that can be compatible with the transferred source hypothesis. 

    \item Exploiting the intrinsic neighborhood structure (NRC) \cite{nrc}: captures the intrinsic structure of the target data by forming clear clusters and encouraging label consistency among data with high local affinity.
    \item Attracting and dispersing (AaD) \cite{aad}: optimizes an objective of prediction consistency by treating SFDA as an unsupervised clustering problem and encouraging local neighborhood features in feature space to have similar predictions.
\end{itemize}

\begin{table*}[]
\centering
\caption{Detailed results of the five SSC cross-domain scenarios in terms of MF1 score.}
\begin{NiceTabular}{l|c|ccccc|c} 
\toprule 
Algorithm & SF & 16$\rightarrow$1 & 9$\rightarrow$14 & 12$\rightarrow$5 & 7$\rightarrow$18 & 0$\rightarrow$11 & AVG\\ \midrule
DDC & \xmark & 55.47$\pm$1.72 & 63.57$\pm$1.43 & 55.43$\pm$2.75 & 67.46$\pm$1.45 & 54.17$\pm$1.79 & 59.22 \\

DCoral &  \xmark& 55.50$\pm$1.74 & 63.50$\pm$1.36 & 55.35$\pm$2.64 & 67.49$\pm$1.50 & 53.76$\pm$1.89 & 59.12 \\

HoMM &\xmark& 55.51$\pm$1.79 & 63.49$\pm$1.14 & 55.46$\pm$2.71 & 67.50$\pm$1.50 & 53.37$\pm$2.47 & 59.06 \\

MMDA &\xmark& 62.92$\pm$0.96 & \underline{71.04$\pm$2.39} & \textbf{65.11$\pm$1.08} & \underline{70.95$\pm$0.82} & 43.23$\pm$4.31 & \underline{62.79} \\

DANN &\xmark& 58.68$\pm$3.29 & 64.29$\pm$1.08 & \underline{64.65$\pm$1.83} & 69.54$\pm$3.00 & 44.13$\pm$5.84 & 60.26 \\

CDAN &\xmark& 59.65$\pm$4.96 & 64.18$\pm$6.37 & 64.43$\pm$1.17 & 67.61$\pm$3.55 & 39.38$\pm$3.28 & 59.04 \\
CoDATS &\xmark& \underline{63.84$\pm$3.36} & 63.51$\pm$6.92 & 52.54$\pm$5.94 & 66.06$\pm$2.48 & 46.28$\pm$5.99 & 58.44 \\
AdvSKM &\xmark& 57.83$\pm$1.42 & 64.76$\pm$3.00 & 55.73$\pm$1.42 & 67.58$\pm$3.64 & 55.19$\pm$4.19 & 60.21 \\ \midrule

 SHOT & \cmark & 59.07$\pm$2.14 & 69.93$\pm$0.46 & 62.11$\pm$1.62 & 69.74$\pm$1.22 &\textbf{ 50.78$\pm$1.90} & 62.33 \\ 
NRC & \cmark & 52.09$\pm$1.89 & 58.52$\pm$0.66 & 59.87$\pm$2.48 & 66.18$\pm$0.25 & \underline{47.55$\pm$1.72} & 56.84 \\ 
AaD & \cmark & 57.04$\pm$2.03 & 65.27$\pm$1.69 & 61.84$\pm$1.74 & 67.35$\pm$1.48 & 44.04$\pm$2.18 & 59.11 \\ \midrule
\textbf{MAPU} & \cmark & \textbf{63.85$\pm$4.63} & \textbf{74.73$\pm$0.64} & 64.08$\pm$2.21 & \textbf{74.21$\pm$0.58} & 43.36$\pm$5.49 & \textbf{64.05} \\
\bottomrule
\end{NiceTabular}
\label{table:eeg}
\end{table*}


\begin{table*}[!ht]
    \centering
    \caption{Detailed results of the five MFD cross-domain scenarios in terms of MF1 score.}
    \begin{NiceTabular}{@{}l|c|ccccc|c@{}} 
    \midrule
    Algorithm &SF & 0$\rightarrow$1 & 1$\rightarrow$0 & 1$\rightarrow$2 & 2$\rightarrow$3 & 3$\rightarrow$1 & AVG \\ \toprule
       DDC & \xmark & 74.50$\pm$5.56 & 48.91$\pm$6.24 & 89.34$\pm$2.16 & 96.34$\pm$3.07 & \textbf{100.0$\pm$0.00} & 81.82  \\ 
        DCoral & \xmark & 79.03$\pm$8.83 & 40.83$\pm$5.01 & 82.71$\pm$0.76 & 98.01$\pm$0.67 & 97.73$\pm$3.93 & 79.66  \\
        HoMM & \xmark & 80.80$\pm$2.46 & 42.31$\pm$5.90 & 84.28$\pm$1.32 & 98.61$\pm$0.08 & 96.28$\pm$6.45 & 80.46  \\
        MMDA & \xmark & 82.44$\pm$4.47 & 49.35$\pm$5.02 & \textbf{94.07$\pm$2.72} & \textbf{100.0$\pm$0.00} & \textbf{100.0$\pm$0.00} & 85.17  \\ 
        DANN & \xmark & 83.44$\pm$1.72 & 51.52$\pm$0.38 & 84.19$\pm$2.10 & \underline{99.95$\pm$0.09} & \textbf{100.0$\pm$0.00} & 83.82  \\ 
        CDAN & \xmark & \underline{84.97$\pm$0.62} & 52.39$\pm$0.49 & 85.96$\pm$0.90 & 99.7$\pm$0.45 & \textbf{100.0$\pm$0.00} & \underline{84.60} \\ 
        CoDATS & \xmark & 67.42$\pm$13.3 & 49.92$\pm$13.7 & \underline{89.05$\pm$4.73} & 99.21$\pm$0.79 & 99.92$\pm$0.14 & 81.10  \\ 
        AdvSKM & \xmark & 76.64$\pm$4.82 & 43.81$\pm$6.29 & 83.10$\pm$2.19 & 98.85$\pm$0.93 & \textbf{100.0$\pm$0.00} & 80.48  \\ \midrule
        
        SHOT & \cmark& 41.99$\pm$2.78 & 57.00$\pm$0.09 & 80.70$\pm$1.49 & 99.48$\pm$0.31 & 99.95$\pm$0.05 & 75.82  \\ 
        NRC & \cmark& 73.99$\pm$1.36 & 74.88$\pm$8.81 & 69.23$\pm$0.75 & 78.04$\pm$11.3 & 71.48$\pm$4.59 & 73.52 \\ 
        AaD & \cmark & 71.72$\pm$3.96 & \underline{75.33$\pm$4.65} & 78.31$\pm$2.26 & 90.07$\pm$7.02 & 87.45$\pm$11.7 & 80.58  \\\midrule 
        \textbf{MAPU}  & \cmark&  \textbf{99.43$\pm$0.51} & \textbf{77.42$\pm$0.16} & 85.78$\pm$7.38 & 99.67$\pm$0.50 & \underline{99.97$\pm$0.05} & \textbf{92.45}\\

\bottomrule
    \end{NiceTabular}
    \label{table:fd}
\end{table*}


\section{Results}
In this section, we rigorously test our approach against state-of-the-art methods in various time series applications. We also assess the versatility of our method by combining it with different SFDA techniques. Furthermore, we compare the effectiveness of our task to other auxiliary tasks on time series data. Lastly, we examine our model's sensitivity to different importance weights and masking ratios. In our MAPU, we leverage SHOT as the base SFDA method. Nevertheless, our approach is not limited to SHOT and can be effectively integrated with other SFDA methods, as demonstrated in our versatility experiments.

\subsection{Quantative Results}
To assess the efficacy of our approach, we evaluate its performance on three different time series datasets, namely, UCIHAR, SSC, and MFD. Tables~\ref{table:har}, \ref{table:eeg}, and \ref{table:fd} present results for five cross-domain scenarios in each dataset, as well as an average performance across all scenarios (AVG). The algorithms are divided into two groups: the traditional UDA methods are marked with \xmark,  while the source-free methods are marked with \cmark.

\subsubsection{Evaluation on UCIHAR Dataset}
The results presented in Table \ref{table:har} show the performance of our MAPU in five cross-subject scenarios. Our method demonstrates superior performance in three of the five scenarios, achieving an overall performance of 89.57\%. This exceeds the second-best source-free method by 3\%. Notably, the source-free methods (i.e., SHOT, NRC, and AaD) perform competitively with conventional unsupervised domain adaptation (UDA) methods that utilize source data. This can be attributed to the two-stage training (i.e., pertaining and adaptation) scheme employed in the source-free methods, which focuses on optimizing the target model for the target domain without considering source performance \cite{cpga}. Furthermore, our MAPU, with its temporal adaptation capability, outperforms all conventional UDA methods, surpassing the best method (i.e., CDAN) by 2.78\%.

\begin{figure*}
     \centering
     \begin{subfigure}[b]{0.65\columnwidth}
         \centering
         \includegraphics[width=\textwidth]{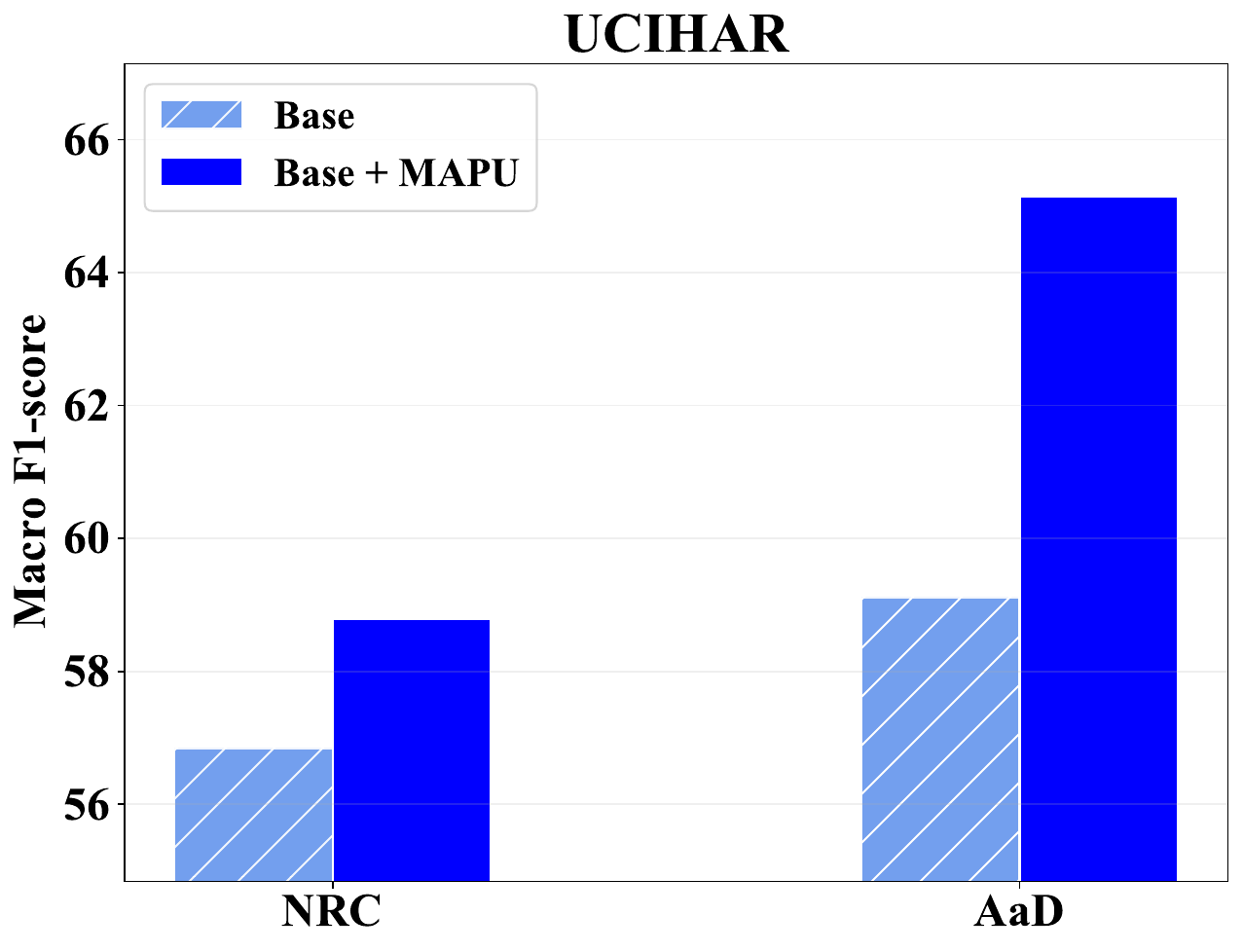}
         \caption{}
         \label{fig:ssc_sens}
     \end{subfigure}
     \hfill
     \begin{subfigure}[b]{0.65\columnwidth}
         \centering
         \includegraphics[width=\textwidth]{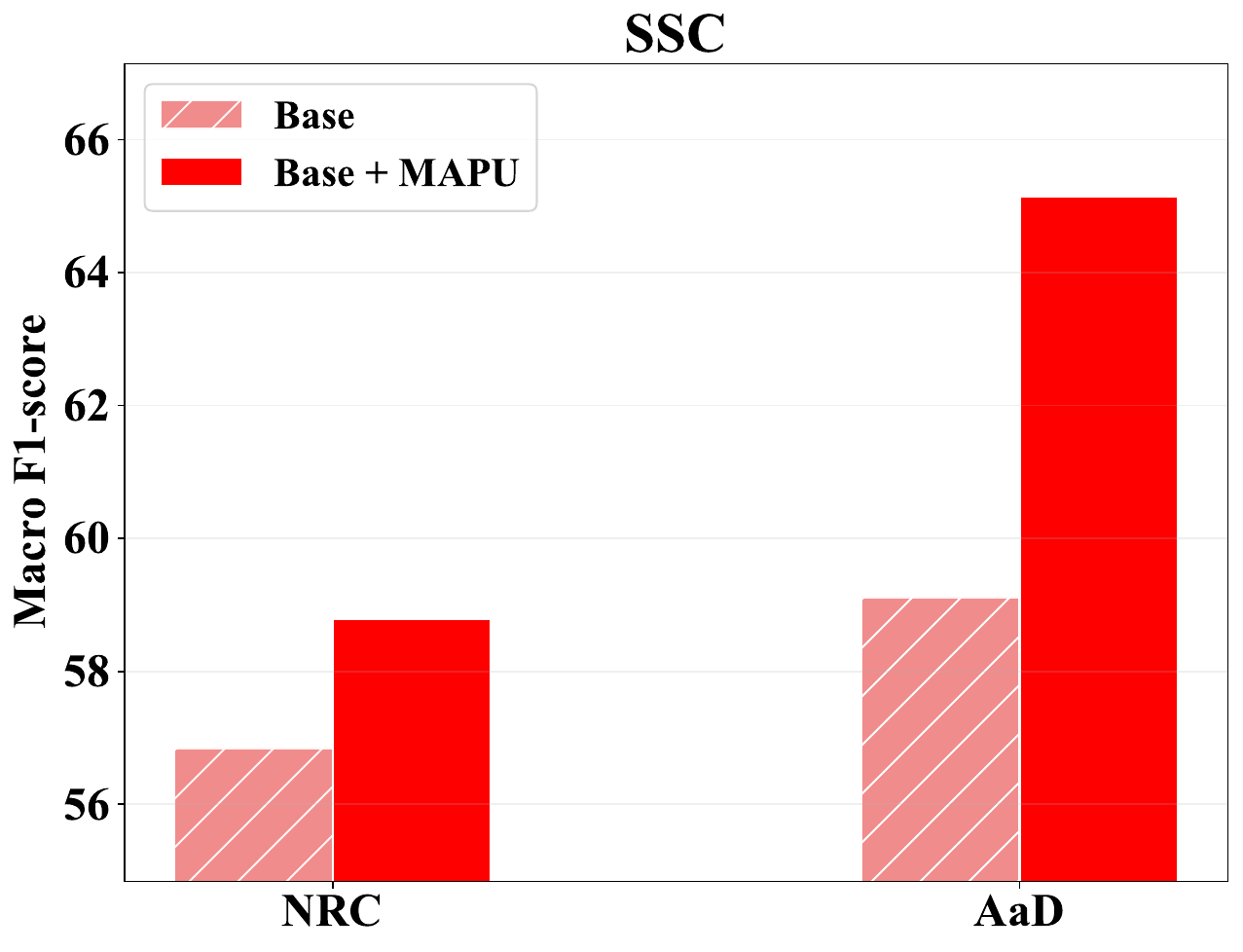}
         \caption{}
         \label{fig:har_sens}
     \end{subfigure}
      \hfill
     \begin{subfigure}[b]{0.65\columnwidth}
         \centering
         \includegraphics[width=\textwidth]{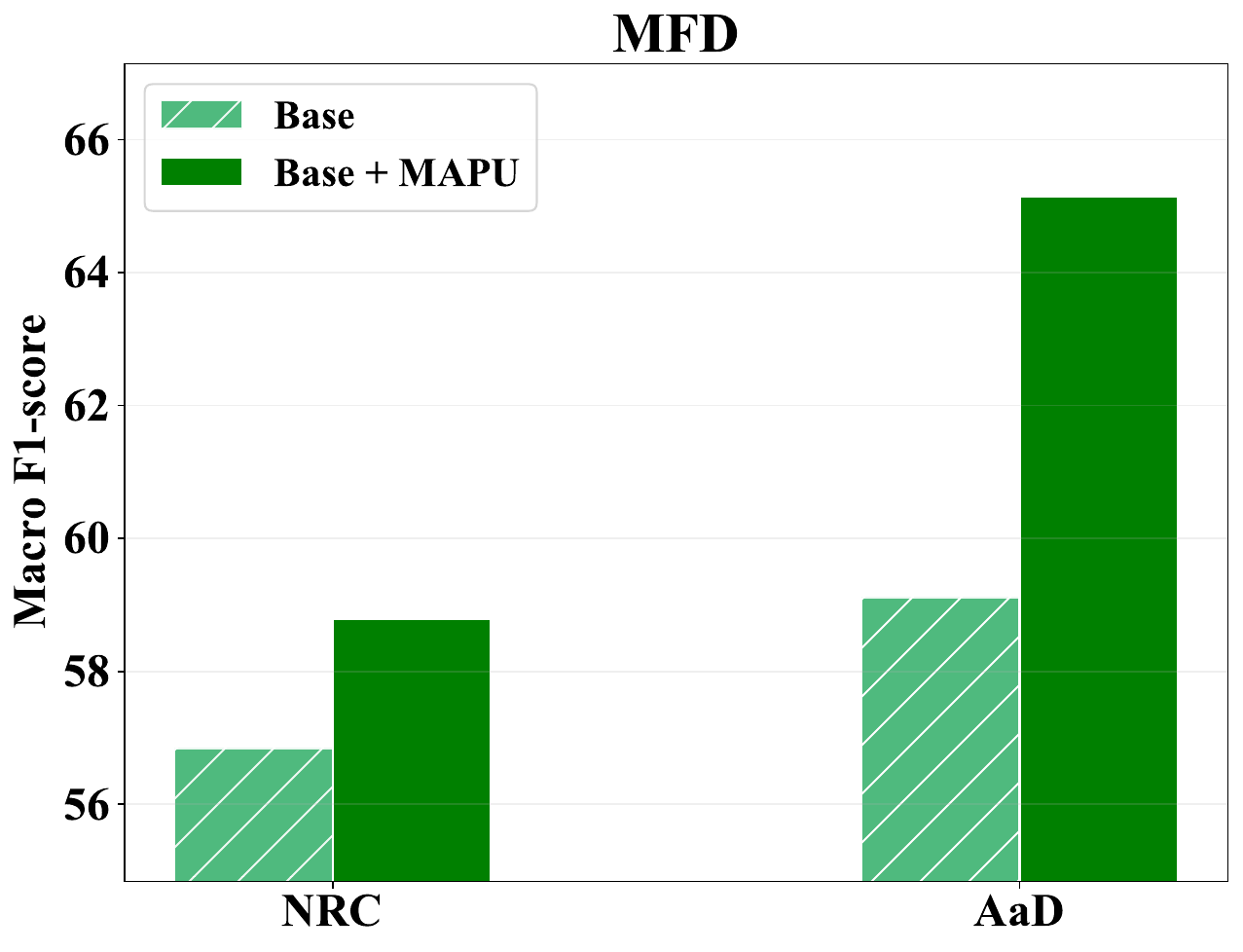}
         \caption{}
         \label{fig:fd_sens}
     \end{subfigure}
     
\caption{Intergrating temporal imputation with existing SFDA methods among the three datasets.}
\label{Fig:universal}
\end{figure*}

\subsubsection{Evaluation on SSC Dataset}
The results of the sleep stage classification task, as presented in Table \ref{table:eeg}, demonstrate the superior performance of our proposed method, MAPU, over other baseline methods. Our MAPU performs best in three out of the five cross-domain scenarios, with an overall performance of 64.05\%. This is higher than the best source-free method, SHOT, and the best conventional UDA method, with an improvement of 1.72\% and 1.27\% respectively. It is worth noting that source-free methods that rely on features clustering, i.e., NRC and AaD, perform poorly on the SSC dataset due to its class-imbalanced nature. However, our MAPU, with its temporal adaptation capability, is able to handle such imbalance and outperform all source-free methods with a maximum improvement of 4.8\% in scenario 16 $\rightarrow$1.

\subsubsection{Evaluation on MFD Dataset}

The results of the Machine Fault Diagnosis (MFD) task, presented in Table \ref{table:fd}, showcase the superior performance of our MAPU when compared to all other baselines. With an average performance of 92.45\%, MAPU exceeds the second-best method by a large margin of 7.85\%. Additionally, MAPU significantly outperforms baseline methods in the hard transfer tasks (i.e., 0$\rightarrow$1 and 1$\rightarrow$0), reaching a 14.46\% improvement in the latter scenario, while performing competitively with other baseline methods in the easy transfer tasks (i.e., $2\rightarrow3$ and $3\rightarrow1$). Compared to source-free methods, our MAPU achieves the best performance in all cross-domains, surpassing the second-best source-free method, AaD, by 11.87\%.

It is worth noting that the performance improvement of our method is relatively large in the MFD dataset compared to other datasets. This is mainly attributed to two reasons. First, the MFD dataset has the longest sequence length among all other datasets, thus, the adaptation of temporal information is more prominent and necessary. Second, unlike other datasets, this dataset has a limited number of classes, i.e., 3 classes, and thus, failing to correctly classify one class can significantly harm the performance.

\begin{table}[]
\caption{Comparing the temporal imputation task with conventional auxiliary tasks for time series adaptation.}
\begin{NiceTabular}{l|ccc}
\toprule
Task & UCIHAR & SSC & MFD  \\ \midrule
SHOT & 86.57 & 62.33 & 75.82   \\ 
SHOT+   Rotation & 86.78 & 60.33 & 84.98  \\
SHOT + Jigsaw & 87.83 & 62.11 & 85.74  \\
\textbf{SHOT + Temporal} & \textbf{89.57} & \textbf{64.05} & \textbf{92.45} \\

\midrule
NRC              & 83.57 & 56.84 & 73.52 \\
NRC+ Rotation    & 71.62 & 56.75 & 72.02 \\
NRC + Jigsaw     & 70.58 & 56.91 & 74.68 \\
\textbf{NRC + Temporal}       & \textbf{86.05} &\textbf{ 58.78} & \textbf{76.34} \\\midrule
AaD              & 84.09 & 59.11 & 80.58 \\
AaD +   Rotation & 71.52 & 59.00 & 84.18 \\
AaD + Jigsaw     & 83.72 & 59.17 & 85.31 \\
\textbf{AaD + Temporal}       & \textbf{87.00} &\textbf{ 64.05} & \textbf{91.11} \\
\bottomrule
\end{NiceTabular}
\label{tab:aux}
\end{table}

\subsection{Ablation Study on Auxiliary Tasks}
To demonstrate the effectiveness of our proposed temporal imputation auxiliary task, we conducted evaluations using various auxiliary tasks, including rotation prediction \cite{SHOT++} and jigsaw puzzle \cite{sl-sfda}. We chose three different SFDA backbones, SHOT, NRC, and AaD, for the auxiliary tasks to eliminate the bias to a specific SFDA method. Table \ref{tab:aux} shows the average performance of five cross-domain scenarios for each dataset. The results show that our temporal imputation task consistently outperforms the other tasks across all datasets, even when combined with different SFDA backbones. Meanwhile, the baseline tasks, including rotation and jigsaw, not only exhibit limited improvement but also consistently harm the performance in many cases across various datasets. This indicates the inadequacy of these tasks for time series data and highlights the importance of considering temporal dynamics to the adaptation performance, as demonstrated by the superior performance of our MAPU approach.

\subsection{Model Analysis}
\subsubsection{Versatility Analysis}
This study investigates the effectiveness of incorporating temporal information into other SFDA methods. To achieve that, we evaluated the performance of three different SFDA methods when used in conjunction with our proposed temporal imputation task on the UCIHAR, SSC, and MFD datasets. Figure \ref{Fig:universal} shows the average performance of five cross-domain scenarios in each dataset. Our results indicate a significant improvement in performance across all tested datasets through the integration of our temporal imputation task. For instance, on the UCIHAR dataset, we saw a notable 3\% boost in performance for the NRC and AaD methods. On the UCIHAR dataset, the NRC and AaD methods all experienced a performance boost of approximately 3\% upon integration with our temporal imputation task. The improvements are consistent across the SSC and MFD datasets, demonstrating our approach's effectiveness in providing temporal adaptation capability to existing SFDA methods that are mainly proposed for visual applications.

\subsubsection{Sensitivity Analysis}
\begin{figure}
    \centering
    \includegraphics[width=0.35\textwidth]{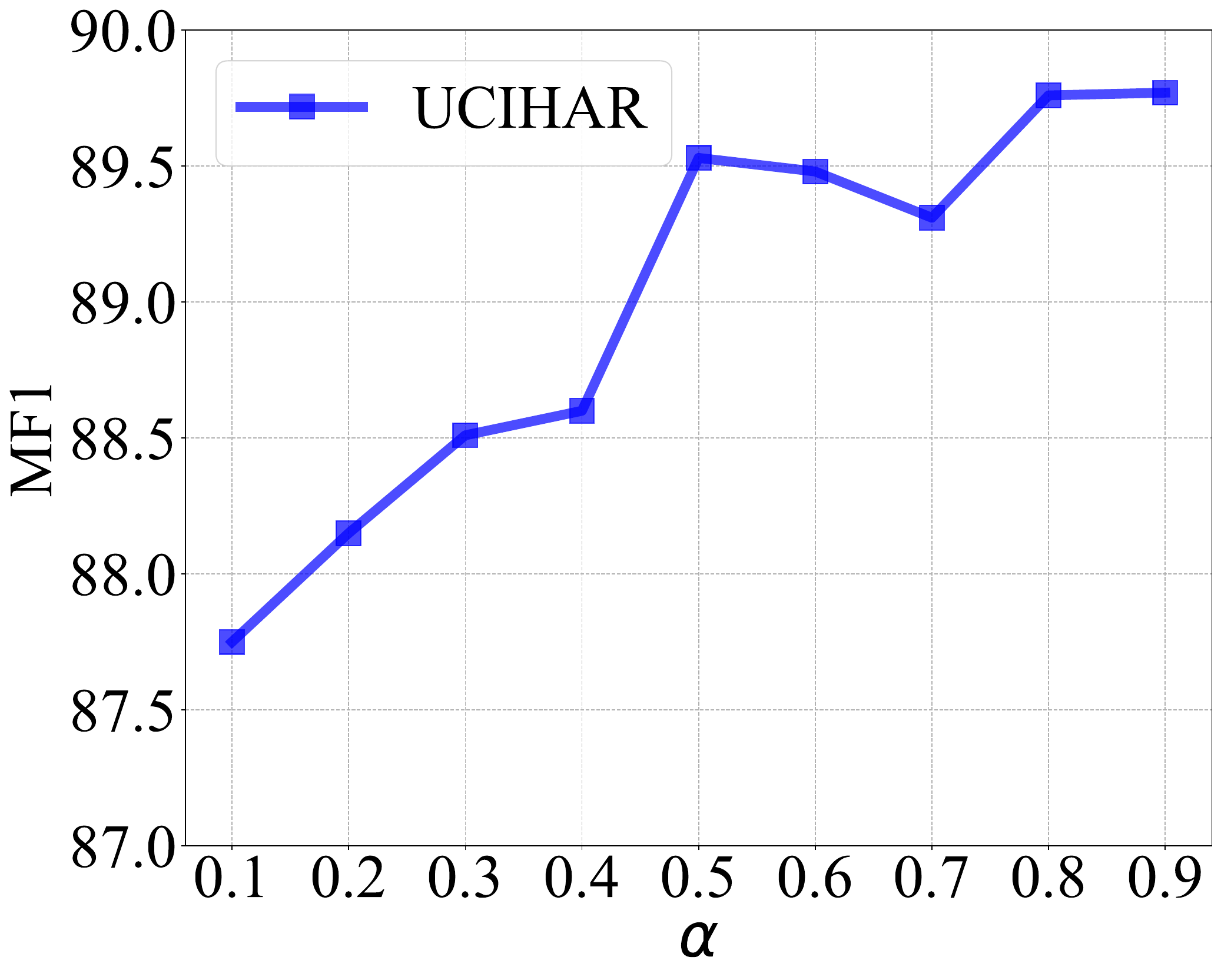}
    \caption{Analysis of adaptation performance with varying relative weight for the temporal imputation component $\alpha$.}
    \label{fig:sens}
\end{figure}

This study evaluates the sensitivity of our temporal imputation component to the relative weight $\alpha$ when integrated with other SFDA methods, as illustrated in Figure \ref{fig:sens}. The results indicate that our model's performance is relatively stable across a range of values for the $\alpha$ parameter. Particularly, the highest MF1 score achieved was 89.77, while the lowest accuracy was 87.75, with a difference of only 2\%. This observed stability may be attributed to the imputation process being carried out on the feature space rather than the input space. As such, the feature space provides a more abstract representation of the data, making the imputation process free of the variations present in the input space. 

\begin{figure}
    \centering
    \includegraphics[width=0.35\textwidth]{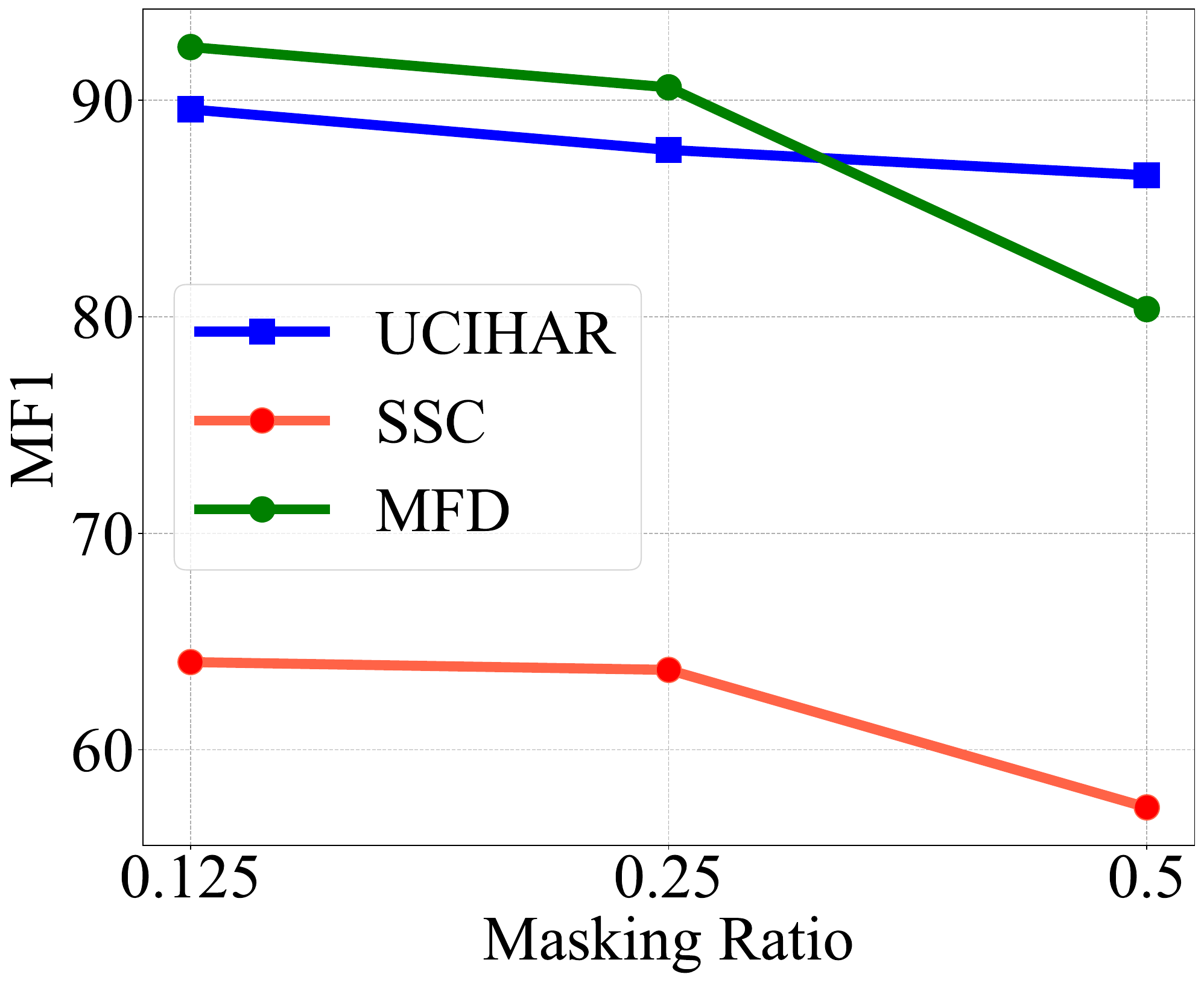}
    \caption{Effect of temporal masking ratio on the adaptation performance for the three datasets.}
    \label{fig:mask}
\end{figure}

\subsubsection{Impact of Masking level}
Here, we systematically examine the impact of the masking ratio on adaptation performance in the context of imputation tasks. Specifically, we employed three different masking ratios (12.5\%, 25\%, and 50\%) and evaluated the performance on the three benchmark datasets. The results, shown in Figure \ref{fig:mask}, reveal a clear trend of improved performance with lower masking ratios. Notably, the best performance was achieved with a masking ratio of 12.5\% across all datasets. These findings suggest that excessive masking may negatively impact the adaptation performance in the imputation task.

\section{Conclusion}
This paper introduced MAsk And imPUte (MAPU), a novel method for source-free domain adaptation on time series data. The proposed method addressed the challenge of temporal consistency in time series data by proposing a temporal imputation task to recover the original signal in the feature space rather than the input space. MAPU is the first method to explicitly account for temporal dependency in a source-free manner for time series data. The effectiveness of MAPU is demonstrated through extensive experiments on three real-world datasets, achieving significant gains over the existing methods. This work highlights the potential of MAPU in addressing the domain-shift problem while preserving data privacy in time series applications.

\section{Acknowledgments}
This work was supported by the Agency of Science Technology and Research under its AME Programmatic (Grant No. A20H6b0151) and its Career Development Award (Grant No. C210112046). 
\balance
\bibliography{refs}
\bibliographystyle{kdd_style/ref_style}

\appendix

\end{document}